 \documentstyle[prl,aps,multicol,epsfig]{revtex}
\begin{document}
\draft \date{\today} \title
  {Irreversible Magnetization of Pin-Free Type II Superconductors}
\author{Ernst Helmut Brandt}
\address{Max Planck Institut f\"ur Metallforschung,
   D-70506 Stuttgart, Germany}
\maketitle

\begin{abstract}
   The magnetization curve of a type II superconductor in general is
hysteretic even when the vortices exhibit no volume or surface pinning.
This geometric irreversibility, caused by an edge barrier for flux
penetration, is absent  only when the superconductor has precisely
ellipsoidal shape or is a wedge with a sharp edge where the flux lines
can penetrate. A quantitative theory of this irreversibility is
presented for pin-free disks and strips with constant thickness.
The resulting magnetization loops are compared with the reversible
magnetization curves of ideal ellipsoids.
\end{abstract}
    \begin{multicols}{2}
    \narrowtext

  The magnetic moment of most superconductors is well known to be
irreversible. After Abrikosov's \cite{1} prediction of quantized
flux lines it became clear \cite{2} that the magnetic hysteresis
is caused by pinning of these vortex lines at inhomogeneities in the
material. Flux-line pinning and the related critical state \cite{3}
were subsequently confirmed quantitatively in numerous
papers \cite{4}. However, similar hysteresis effects were also
observed \cite{5} in type I superconductors, which do not contain
flux lines but normal conducting domains, and in type II
superconductors with negligible pinning. In these two cases the
magnetic irreversibility is caused by a geometric (specimen-shape
dependent) barrier which delays the penetration of magnetic flux
but not its exit. In this respect the geometric barrier behaves
similar to the Bean-Livingston barrier \cite{6,7} for vortices
penetrating a parallel surface.

  The geometric irreversibility is most pronounced for thin films
of constant thickness in a perpendicular field. It is absent
only when the superconductor is of exactly ellipsoidal shape or
is tapered like a wedge with a sharp edge where flux penetration is
facilitated. In ellipsoids the inward directed driving force exerted
on the vortex ends by the surface screening currents is exactly
compensated by the vortex line tension \cite{8}, and thus the
magnetization is reversible. In specimens with constant thickness
(i.e.\ rectangular cross-section) this line tension
opposes the penetration of flux lines at the four corner lines,
thus causing an edge barrier; but as soon as two penetrating vortex
segments join at the equator they contract and are driven to
the specimen center by the surface currents, see Fig.\ 1 below. As
opposed to this, when the specimen profile is tapered and has a sharp
edge, the driving force even in very weak applied field exceeds the
restoring force of the line tension such that there is no edge
barrier. The resulting absence of hysteresis in wedge-shaped
samples was nicely shown by Morozov et al.\ \cite{9}.

   An elegant analytical theory of the field and current profiles in
thin superconductor strips with an edge barrier has been presented
by Zeldov et al.\ \cite{10}, see also the extensions \cite{11}.
With increasing applied field $H_a$, the magnetic flux does not
penetrate until an entry field $H_{\rm en}$ is reached;
at $H_a = H_{\rm en}$ the flux immediately jumps to
the center, from where it gradually fills the entire strip or
disk. This behavior in increasing $H_a$ is similar to that of
thin films with artificially enhanced pinning near the
edge \cite{11,12}, but in decreasing $H_a$ the behavior is
different: In films with enhanced edge pinning (critical current
density $J_{c,{\rm edge}}$) the current density $J$ at the edge
immediately jumps from $+J_{c,{\rm edge}}$ to  $-J_{c,{\rm edge}}$
when the ramp rate inverses sign, while in pin-free
films with geometric barrier the current density at the edge
first stays constant or even increases and then
gradually decreases and reaches zero at $H_a=0$.
The entry field $H_{\rm en}$ was estimated for pin-free thin strips
in Refs.\ \onlinecite{10,13}, see also Refs.\ \onlinecite{14,15}.

  In this letter the geometry-caused magnetic irreversibility of
ideal pin-free type II superconductors is calculated and discussed
for the two most important examples of circular disks (or cylinders)
and long strips (or slabs) with rectangular profile of arbitrary
aspect ratio $b/a$. I present flux-density profiles and magnetization
loops and give explicit expressions for the entry field $H_{\rm en}$
and for the reversibility field $H_{\rm rev}$ above which the
magnetization curve is reversible. Finally, the modification of
these results by volume pinning is briefly mentioned.

   Let us first consider the magnetization of ideal
ellipsoids. If the superconductor is homogeneous and isotropic,
the magnetization curves $M(H_a; N)$ are reversible and may be
characterized by a demagnetizing factor $N$ with $0 \le N \le 1$.
If $H_a$ is along one of the three principal axes of the
ellipsoid then $N$ is a scalar. One has $N=0$ for long specimens
in parallel field, $N=1$ for thin films in perpendicular field, and
$N=1/3$ for spheres. If the magnetization curve in parallel field is
known, $M(H_a; 0) = B/\mu_0 -H_a$ where $B$ is the flux density or
induction inside the ellipsoid, then the homogeneous
magnetization of the general ellipsoid, $M(H_a; N)$, follows
from the implicit equation
  \begin{eqnarray}    
    H_i = H_a - N\, M(H_i; 0)\,.
  \end{eqnarray}
Solving Eq.\ (1) for the effective internal field $H_i$, one
obtains $M=M(H_a; N) = M(H_i; 0)$. In particular, for
the Meissner state ($B \equiv 0$) one finds $M(H_a;0) = -H_a$ and
    \begin{eqnarray}    
    M(H_a; N) = - {H_a \over 1-N} ~~~{\rm for}~~
               |H_a| \le (1-N)H_{c1} \,.
  \end{eqnarray}
At the lower critical field $H_{c1}$ one has $H_i = H_{c1}$,
$H_a = H_{c1}' = (1-N) H_{c1}$, $B=0$, and $M=-H_{c1}$. Near the
upper critical field $H_{c2}$ one has an approximately linear
$M(H_a; 0) =  \gamma (H_a -H_{c2}) <0$ with $\gamma >0$,
yielding
  \begin{eqnarray}    
    M(H_a; N) = {\gamma \over 1 + \gamma N } (H_a - H_{c2})
   ~~~{\rm for}~~ H_a \approx H_{c2} \,.
  \end{eqnarray}
Thus, if the slope $\gamma \ll 1$ is small (and in general, if
$|M/H_a| \ll 1$ is small), demagnetization effects may be
disregarded and one has $M(H_a; N) \approx M(H_a; 0)$.

   The ideal magnetization curve of type II superconductors with
$N=0$, $M(H_a; 0)$ or $B(H_a;0) =H_a +M(H_a;0)$, may be calculated
from Ginzburg-Landau (GL) theory \cite{16}, but any other model curve
may be used provided $M(H_a; 0) = -M(-H_a; 0)$ has a vertical slope
at $H_a =H_{c1}$ and decreases monotonically in size for $H_a >H_{c1}$.
For simplicity in this letter I shall assume $H_{c1} \ll H_{c2}$
(i.e.\ large GL parameter $\kappa \gg 1$) and $H_a \ll H_{c2}$.
To illustrate the essential features I may thus use the realistic model
$M(H_a; 0) = - H_a$  for $|H_a| \le H_{c1}$ and
   \epsfxsize= .70\hsize   \vskip 1.0\baselineskip
\centerline{ \epsffile{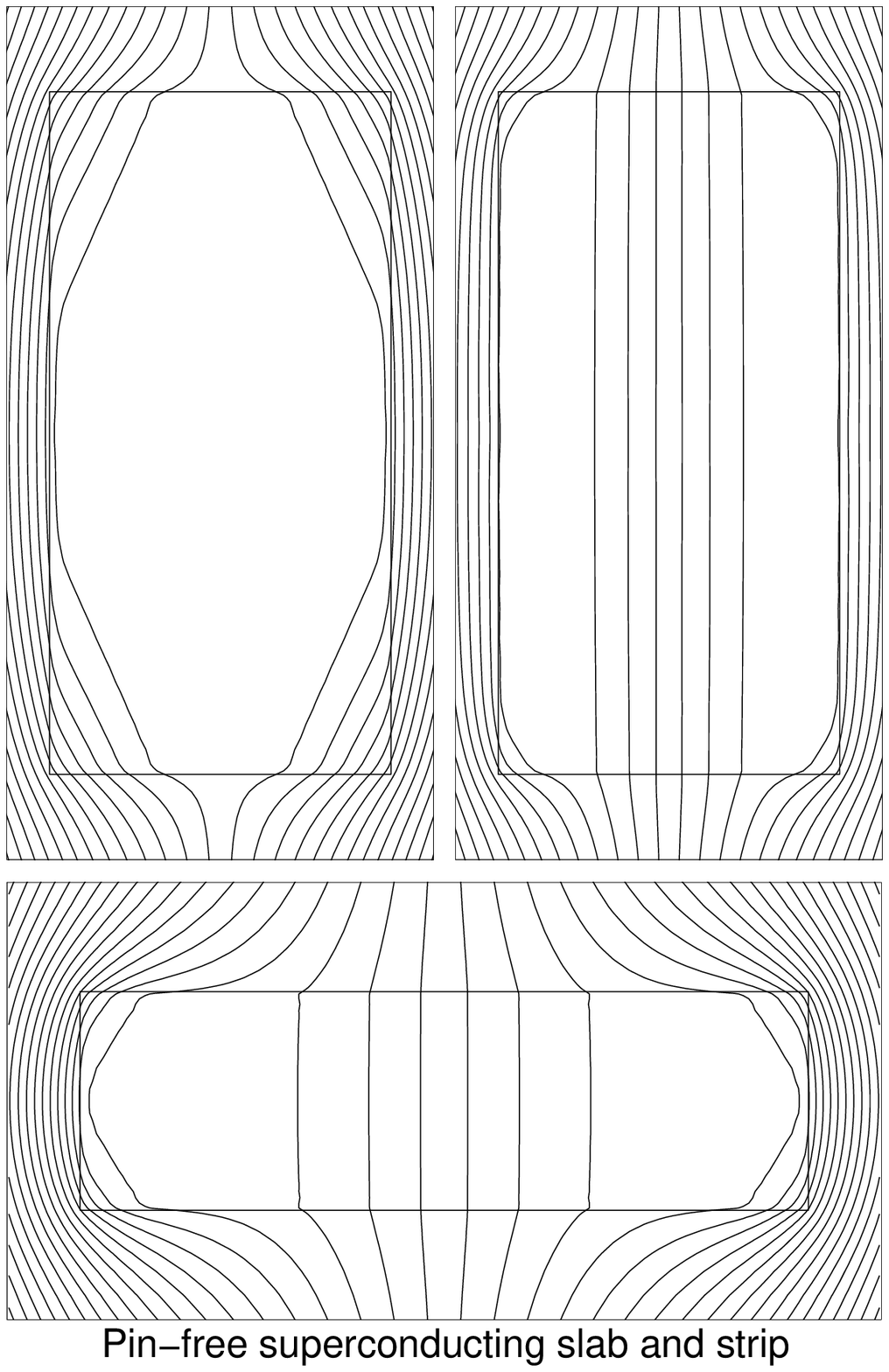}}
 \begin{figure}[F1]
\caption{The magnetic field lines of $B(x,y)$ in slabs or strips with
 aspect ratio $b/a=2$ (top) and $b/a =0.3$ (bottom) in perpendicular
 magnetic field $H_a$.  Top left: $H_a/H_{c1} = 0.66$,
 in increasing field shortly below the entry field
 $H_{\rm en}/H_{c1}=0.665$. Top right: $H_a/H_{c1} = 0.5$, decreasing
 field.  Bottom: $H_a/H_{c1} = 0.34$ in increasing field just
 above $H_{\rm en}/H_{c1}= 0.32$. The field lines of cylinders look
 very similar. Note the straight field lines in the corners,
 corresponding to flux lines under tension.
    } \end{figure}
  \begin{eqnarray}    
    M(H_a; 0) = (H_a/|H_a|) (|H_a|^3 - H_{c1}^3 )^{1/3} - H_a
  \end{eqnarray}
for $|H_a| > H_{c1}$, see the curve labeled $\infty$ in Fig.\ 3 below.

In nonellipsoidal superconductors the induction ${\bf B(r})$ in
general is not homogeneous, and so the concept of a demagnetizing
factor does not work. However, when the magnetic moment
${\bf m} = {1\over 2} \int {\bf r \times J(r)} d^3 r$ is directed
along $H_a$, one may define an {\it effective demagnetizing factor}
$N$ which in the Meissner state ($B\equiv 0$) yields the same
slope $M/H_a = -1/(1-N)$, Eq.\ (2), as an ellipsoid with the same
volume $V$. Here the definition $M=m/V$ with $m= {\bf m H}_a/H_a$
is used.  For long strips and circular disks or cylinders with
cross-section $2a \times 2b$ in a perpendicular or axial magnetic
field along the thickness $2b$, approximate expressions for the
slopes $M/H_a = m/(VH_a)$ are given in Refs.\ \onlinecite{17,18}.
Using this and defining $q \equiv (|M/H_a| -1)(b/a)$, one obtains
the effective $N$ for any aspect ratio $b/a$ in the form
  \begin{eqnarray}    
  N &=& 1 - 1/(1 + q a/b) \,, \nonumber \\
  q_{\rm strip} &=&  {\pi \over 4 } +0.64
   \tanh\Big[ 0.64{b\over a} \ln \Big( 1.7+1.2{a\over b} \Big) \Big]
    \,, \nonumber \\
  q_{\rm disk}  &=&  {4 \over 3\pi } +{2 \over 3\pi }
   \tanh\Big[ 1.27{b\over a} \ln \Big(1 +{a\over b} \Big) \Big] \,.
       \end{eqnarray}
In the limits $b \ll a$ and $b \gg a$, formulae (5) are exact,
and for general $b/a$ the relative error is $< 1\%$.
For $a=b$ (square cross-section) they yield for the strip $N = 0.538$
(while $N=1/2$ for a circular cylinder in perpendicular field) and for
the short cylinder  $N = 0.365$ (while $N=1/3$ for the sphere).

    Next we consider the full, irreversible magnetization curves
$M(H_a)$ of pin-free strips and cylinders with cross section
$2a \times 2b$. Appropriate continuum equations and algorithms
(which apply also to pinning) have been proposed recently by Labusch
and Doyle \cite{19} and by the author \cite{20}, based on the
Maxwell equations and on constitutive laws which describe flux flow
and pinning [or thermal depinning expressed, e.g., by an electric
field ${\bf E(J,B)}$] and the reversible magnetization in absence
of pinning, $M(H_a;0)$. Here I shall use the method [20] and
the model $M(H_a;0)$, Eq.\ (4). The pin-free flux dynamics
will be described as viscous motion by
${\bf E}=\rho_{\rm FF}(B) {\bf J}$  with flux-flow resistivity
$\rho_{\rm FF} \propto B$. In both methods the $M(H_a;0)$ law
enters the driving force density on the vortices,
${\bf J_H} \times {\bf B}$ with definition
${\bf J_H} = \nabla \times {\bf H}$, where ${\bf H}({\bf B})$
is obtained by inverting the relation
${\bf B(H}) = {\bf H} + {\bf M(H};0)$.

   While method [19] considers a magnetic charge density on the
specimen surface which causes an effective field ${\bf H}_i({\bf r})$
inside the superconductor, our method [20] couples the
arbitrarily shaped superconductor to the external field
${\bf B(r},t)$ via surface screening currents: In a first step
the vector potential ${\bf A(r},t)$ is calculated for given
current density ${\bf J}$; then this relation (a matrix) is
   \epsfxsize= .85\hsize  \vskip 0.5\baselineskip
\centerline{ \epsffile{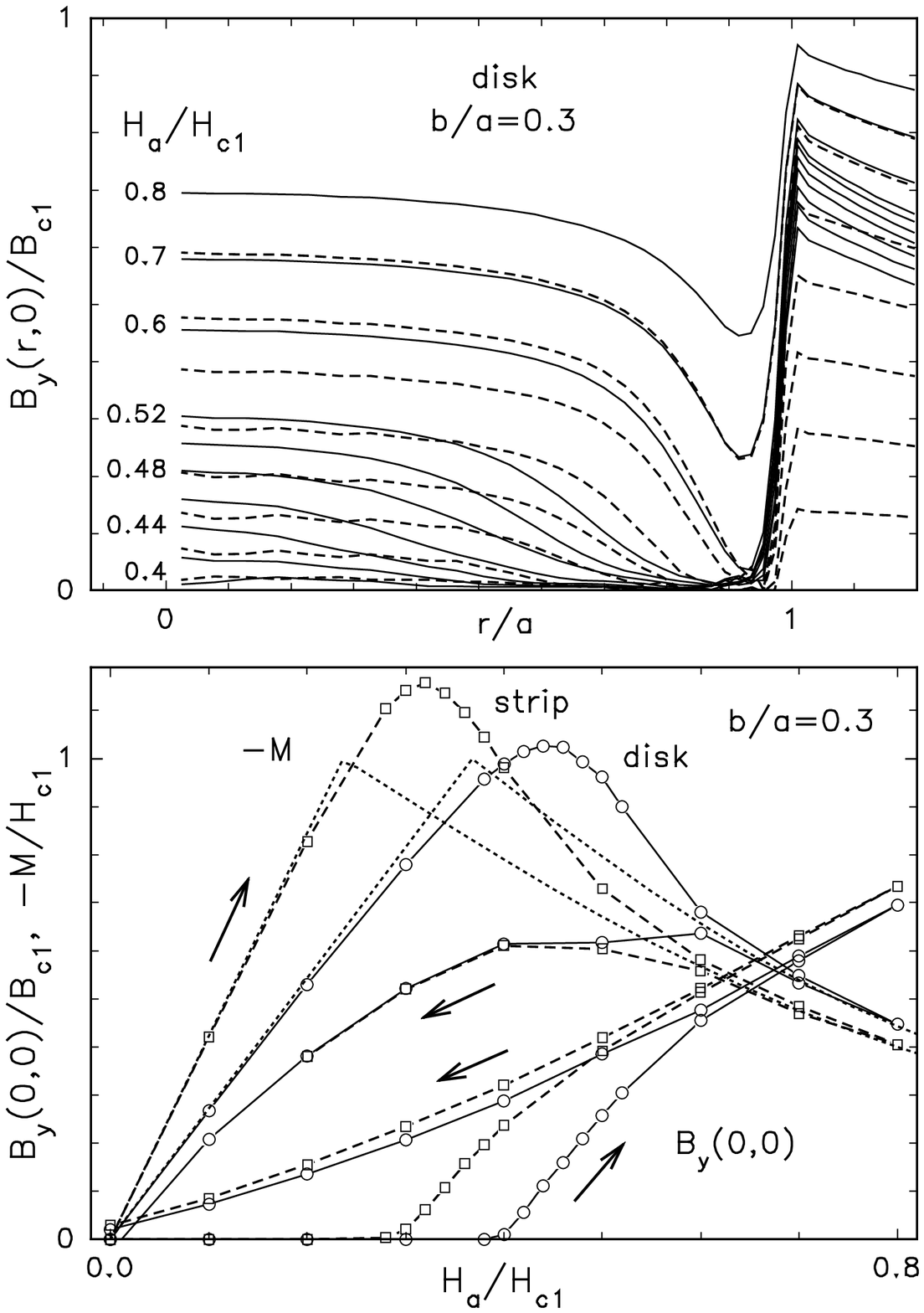}}
 \begin{figure}[F2]
\caption{Top: Profiles of the axial magnetic induction $B_y(r,y)$ in
the midplane $y=0$ of a pin-free superconductor disk with aspect ratio
 $b/a=0.3$ in increasing field (solid lines) and then decreasing
field (dashed lines), plotted
 at $H_a/H_{c1}$ = 0.4, 0.42, $\dots$, 0.5, 0.52, 0.6, 0.5,
 0.7, 0.8, 0.7, 0.6, $\dots$, 0.1, 0. $B_{c1}=\mu_0 H_{c1}$.
 Bottom: The induction $B_y(0,0)$ in the center of the same disk
 (solid line) and of a strip (dashed line), both with $b/a=0.3$.
 The symbols mark the field values at which the profiles are
 taken. Also shown is the magnetization loop for the same
 disk and strip and the corresponding reversible magnetization
 (dotted lines), see also Fig.\ 3.
  } \end{figure}
\noindent
inverted to obtain ${\bf J}$ for given ${\bf A}$
and given ${\bf H}_a$;
next the induction law is used to obtain the electric field [in our
symmetric geometry one has
${\bf E(J,B)} = -\partial {\bf A}/\partial t$\,], and finally the
constitutive law ${\bf E=E(J,B)}$ is used to eliminate
${\bf A}$ and ${\bf E}$ and
obtain one single integral equation for ${\bf J(r},t)$ as
a function of ${\bf H}_a(t)$, without having to compute
${\bf B(r},t)$ outside the specimen. This method in general
is fast and elegant; but so far the algorithm is restricted
to  moderate aspect ratios, $0.03 \le b/a \le 30$, and to a number
of grid points not exceeding 1000 (on a Personal Computer).
Improved accuracy is expected by combining methods (19)
(working best for small $b/a$) and (20).

   The penetration and exit of flux computed by method [20] is
illustrated in Figs.\ 1 and 2 for isotropic strips and disks without
volume pinning, using a flux-flow resistivity
$\rho_{\rm FF} = \rho B({\bf r})$ with $\rho = 140$ (strip) or
$\rho = 70$ (disk) in units where $H_{c1} =a = \mu_0 =|dH_a/dt| =1$.
The profiles of the induction $B_y(r,y)$ taken along the midplane
$y=0$ of the thick disk in Fig.\ 2 have a pronounced minimum
\epsfxsize= .95\hsize  \vskip 0.5\baselineskip
\centerline{ \epsffile{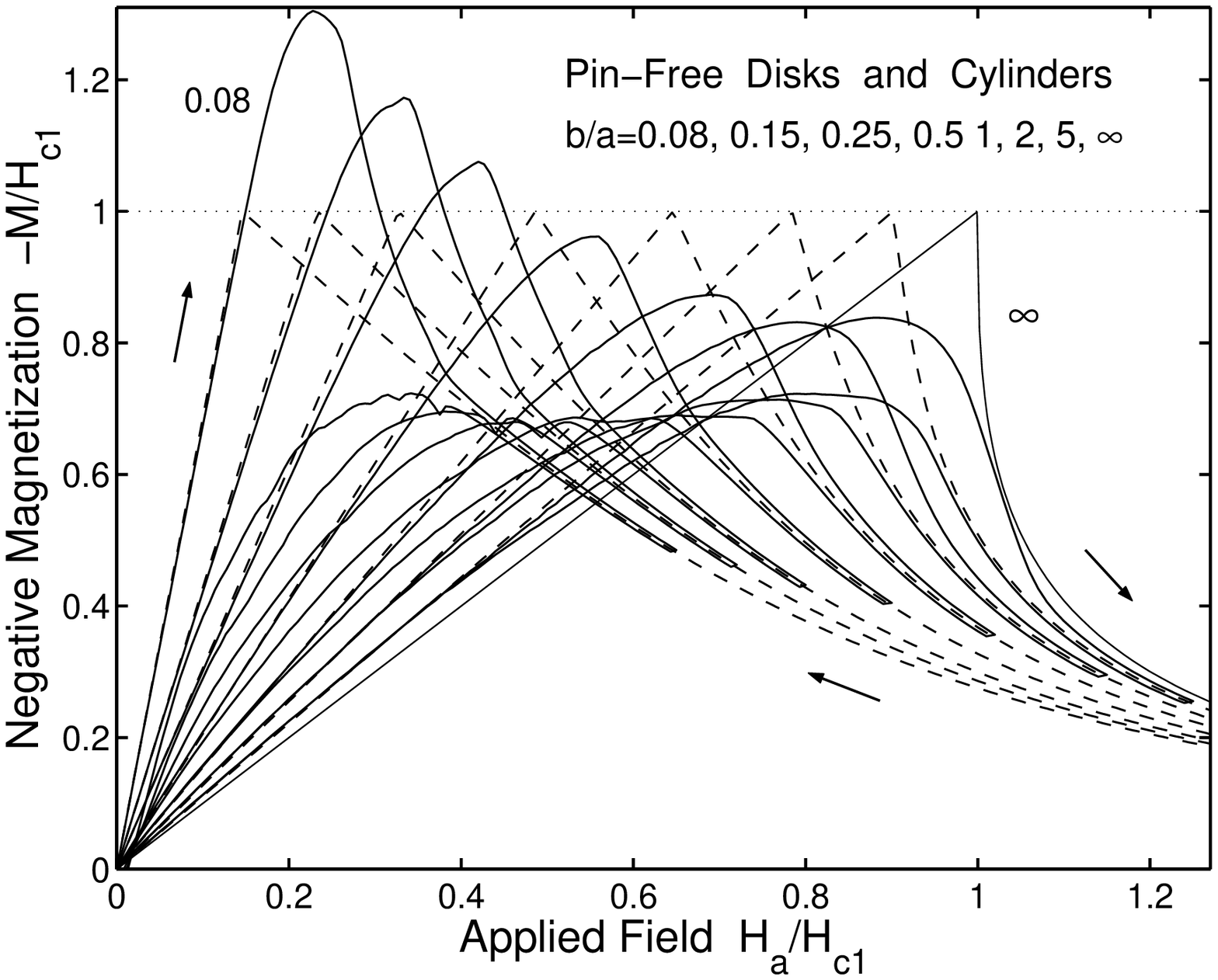}}
 \begin{figure}[F3]
\caption{Irreversible magnetization curves $-M(H_a)$ of
 pin-free circular disks or cylinders with aspect ratios
 $b/a=$ 0.08, 0.15, 0.25, 0.5, 1, 2, 5, and $\infty$ in axial field
 (solid lines). In these type II superconductors the irreversibility
 is due to a purely geometric edge barrier for flux penetration.
  The dashed curves are the reversible magnetization curves of the
  corresponding ellipsoid defined by Eqs.\ (1,4,5).
  } \end{figure}
\noindent
near the edge $r=a$, precisely in the region where strong screening
currents flow. Away from the  edges, the current density
${\bf J} = \nabla \times {\bf B}/\mu_0$ is nearly zero; note
the parallel field lines in Fig.\ 1. The quantity
${\bf J_H} = \nabla\times{\bf H(B)}$  which enters the Lorentz
force density ${\bf J_H \times B}$, is even exactly zero since
we assume absence of pinning. Our finite flux-flow parameter $\rho$
and finite ramp rate $dH_a / dt = \pm 1$ mean a dragging force
which, similar to pinning, causes a weak hysteresis and a small
remanent flux at $H_a =0$; this effect may be reduced by choosing
larger resistivity and slower ramping.

   The induction $B_y(0,0)$ in the specimen center in Fig.\ 2
performs a hysteresis loop very similar to the magnetization loops
$M(H_a)$ shown in Figs.\ 2, 3. Both loops are symmetric, e.g.,
$M(-H_a) = -M(H_a)$. The maximum of $M(H_a)$ defines a field of
first flux entry $H_{\rm en}$, which closely coincides with the
field $H'_{\rm en}$ at which $B_y(0,0)$ starts to appear. The
computed entry fields are well fitted by
       \begin{eqnarray}    
  H_{\rm en}^{\rm strip}/H_{c1}  &=& \tanh \sqrt{0.36 b/a} \,,
                             \nonumber \\
  H_{\rm en}^{\rm disk}/H_{c1}   &=& \tanh \sqrt{0.67 b/a} \,.
       \end{eqnarray}
These formulae are good approximations for all aspect ratios
$0 < b/a < \infty$, see also the estimates of
$H_{\rm en} \approx \sqrt{b/a}$ for thin strips in
Refs.\ \onlinecite{10,13}.

   The virgin curve of the irreversible $M(H_a)$ of strips and
disks at small $H_a$ coincides with the ideal Meissner straight
line $M = -H_a/(1-N)$ of the corresponding ellipsoid, Eqs.\ (2,5).
When the increasing $H_a$ approaches $H_{\rm en}$, flux starts to
penetrate into the corners in form of
\epsfxsize= .95\hsize  \vskip 0.5\baselineskip
\centerline{ \epsffile{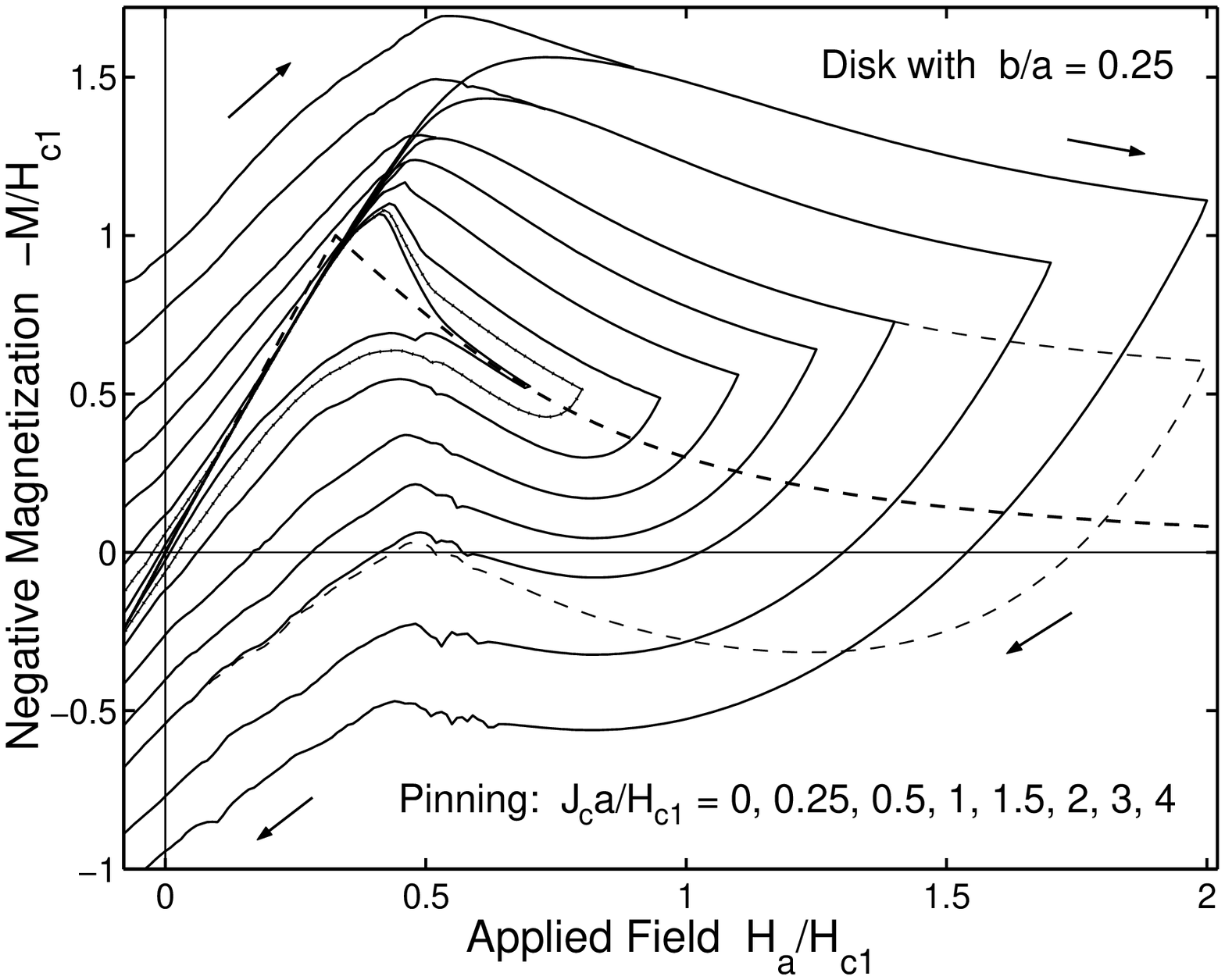}}
 \begin{figure}[F4]
\caption{Magnetization curves of a thick disk with aspect ratio
 $b/a=0.25$  for various degrees of volume pinning,
 $J_c$ = 0, 0.25, 0.5, 1, 1.5, 2, 3, 4 in units $H_{c1}/a$, and
 for various sweep amplitudes.
 The inner loop belongs to the pin-free disk ($J_c=0$), the outer
 loop to strongest pinning.  Also shown is the reversible
 magnetization curve of the corresponding ellipsoid (dashed curve).
 All loops are symmetric, $M(-H_a)= -M(H_a)$.
  } \end{figure}
\noindent
stretched flux lines (Fig.\ 1)
and thus $|M(H_a)|$ falls below the Meissner line.
At $H_a= H_{\rm en}$ flux penetrates and jumps
to the center, and $|M(H_a)|$ starts to decrease. In decreasing
$H_a$, this barrier is absent. As can be seen in Fig.\ 3, above
some field $H_{\rm rev}$, the magnetization curve $M(H_a)$ becomes
reversible and exactly coincides with the curve of the ellipsoid
defined by Eqs.\ (1, 4, 5) (in the quasistatic limit with
$\rho^{-1} dH_a/dt \to 0$). The irreversibility field $H_{\rm rev}$
is difficult to compute since, in our present algorithm, it
slightly depends on the choices of the flux-flow parameter
$\rho$ (or ramp rate) and of the numerical grid, and also on
the model for $M(H_a; 0)$. In the interval
$0.08 \le b/a \le 5$ we find with relative error of $3\%$,
       \begin{eqnarray}    
  H_{\rm rev}^{\rm strip}/H_{c1}  &=& 0.65 +0.12 \ln{(b/a)} \,,
                              \nonumber \\
  H_{\rm rev}^{\rm disk}/H_{\rm c1}   &=& 0.75 +0.15 \ln{(b/a)} \,.
       \end{eqnarray}
This fit obviously does not apply to $b/a \ll 1$ (since $H_{\rm rev}$
should exceed $H_{\rm en} > 0$) nor to $b/a \gg 1$ (where $H_{\rm rev}$
should be close to $H_{\rm c1}$). The limiting value of
$H_{\rm rev}$ for thin films with $b \ll a$ is thus not known
at present.

   Remarkably, the irreversible magnetization curves $M(H_a)$ of
pin-free strips and disks fall on top of each other if the strip is
chosen twice as thick as the disk,
$(b/a)_{\rm strip} \approx 2(b/a)_{\rm disk}$. This striking
coincidence holds for all aspect ratios $0 < b/a < \infty$ and
can be seen from each of Eqs.\ (5-7): The effective $N$ [or virgin
slope $1/(1-N)$], the entry field $H_{\rm en}$, and the reversibility
field $H_{\rm rev}$  are nearly equal for strips and disks with half
thickness, or for slabs and cylinders with half length.

   Another interesting feature of the pin-free magnetization loops is
that the maximum of $|M(H_a)|$ exceeds the maximum of the reversible
curve (equal to $H_{c1}$) when $b/a \le 0.8$ for strips and
$b/a \le 0.4$ for disks, but at larger $b/a$ it falls below $H_{c1}$.
The maximum magnetization may be estimated from the slope of the
virgin curve $1/(1-N)$, Eq.\ (5), and from the field of first
flux entry, Eq.\ (6).

    Finally, Fig.\ 4 shows how the irreversible magnetization loop
is modified when volume pinning of the flux lines is switched on.
Increasing critical current density $J_c$ (in natural units
$H_{c1}/a$) inflates the loops nearly symmetrically about the
pin-free loop or (above $H_{\rm rev}$) about the reversible
curve, and the maximum of $|M(H_a)|$ shifts to higher fields.
Above $H_{\rm rev}$ the width of the loop is nearly proportional
to $J_c$, as expected from previous theories \cite{4,17,18} which
assumed $H_{c1}=0$, but at small fields the influence of finite
$H_{c1}$ is clearly seen up to rather strong pinning.

\end{multicols}

\vspace{-0.3 cm}
\references
         \vspace{-1.8 cm}

\bibitem{1} A.\ A.\ Abrikosov, Zh.\ Eksp.\ Teor.\ Fiz.\ {\bf 32},
            1442 (1957) [Sov.\ Phys.-JETP {\bf 20}, 480 (1965)].
\bibitem{2} P.\ W.\ Anderson, \prl{\bf 9}, 309 (1962).
\bibitem{3} C.\ P.\ Bean, Rev.\ Mod.\ Phys.\ {\bf 36}, 31 (1964).
\bibitem{4} A.\ M.\ Campbell and J.\ E.\ Evetts, Adv.\ Phys.\
            {\bf 72}, 199 (1972).

\bibitem{5} J.\ Provost, E.\ Paumier, and A.\ Fortini,
            J.\ Phys.\ F {\bf 4}, 439 (1974); A.\ Fortini,
            A.\ Haire, and E.\ Paumier, \prb{\bf 21}, 5065 (1980).
\bibitem{6} C.\ P.\ Bean and J.\ D.\ Livingston, \prl{\bf 12},
            14 (1964).
\bibitem{7} L.\ Burlachkov, \prb{\bf 47}, 8056 (1993).

\bibitem{8} M.\ V.\ Indenbom, H.\ Kronm\"uller, T.\ W.\ Li,
            P.\ H.\ Kes, and A.\ A.\ Menovsky,
            Physica C {\bf 222}, 203 (1994);    
            M.\ V.\ Indenbom and E.\ H.\ Brandt,
            \prl{\bf 73}, 1731 (1994);
            E.\ H.\ Brandt, Rep.\ Prog.\ Phys.\ {\bf 58}, 1465 (1995).

\bibitem{9} N.\ Morozov et al., Physica C {\bf 291}, 113 (1997).

\bibitem{10} E.\ Zeldov, A.\ I.\ Larkin, V.\ B.\ Geshkenbein,
       M.\ Konczykowski, D.\ Majer, B.\ Khaykovich, V.\ M.\ Vinokur,
       and H.\ Strikhman, \prl{\bf 73}, 1428 (1994).
\bibitem{11} E.\ Zeldov et al., Physica C {\bf 235-240}, 2761 (1994);
       B.\ Khaykovich et al., Physica C {\bf 235-240}, 2757 (1994);
       N.\ Morozov et al., \prl{\bf 76}, 138 (1996).

\bibitem{12} Th.\ Schuster, M.\ V.\ Indenbom, H.\ Kuhn, E.\ H.\ Brandt,
       and M.\ Konczykowski, \prl{\bf 73}, 1424 (1994).

\bibitem{13} M.\ Benkraouda and J.\ R.\ Clem, \prb{\bf 53}, 5716 (1996);
       \prb{\bf 58}, 15103 (1998).

\bibitem{14} I.\ L.\ Maksimov and A.\ A.\ Elistratov,
       Pis'ma Zh.\ Eksp.\ Teor.\ Fiz. {\bf 61}, 204 (1995)
       [Sov.\ Phys.\ JETP Lett. {\bf 61}, 208 (1995)].

\bibitem{15} A.\ V.\ Kuznetsov, D.\ V.\ Eremenko, and V.\ N.\ Trofimov,
        \prb{\bf 56}, 9064 (1997); \prb{\bf 57}, 5412 (1998).

\bibitem{16} E.\ H.\ Brandt, \prl{\bf 78}, 2208 (1997).  

\bibitem{17} E.\ H.\ Brandt, \prb{\bf 54}, 4246 (1996). 

\bibitem{18} E.\ H.\ Brandt, \prb{\bf 58}, 6506, 6523 (1998).

\bibitem{19} R.\ Labusch and T.\ B.\ Doyle, Physica C {\bf 290}, 143
       (1997); T.\ B.\ Doyle, R.\ Labusch, and R.\ A.\ Doyle,
       Physica C {\bf 290}, 148 (1997).

\bibitem{20} E.\ H.\ Brandt, \prb{\bf 59}, 3369 (1999).
\end{document}